\documentclass[conference]{IEEEtran}
\IEEEoverridecommandlockouts
\usepackage{cite}
\usepackage{amsmath,amssymb,amsfonts}
\usepackage{algorithmic}
\usepackage{graphicx}
\usepackage{textcomp}
\usepackage{xcolor}
\usepackage{amsmath}
\usepackage{graphicx}
\usepackage{multirow}
\usepackage[none]{hyphenat}
\usepackage{float}
\usepackage{subfig}
\usepackage{dblfloatfix}
\usepackage{t1enc}
\usepackage{times}
\usepackage[usestackEOL]{stackengine}
\usepackage[numbers,sort&compress]{natbib}

\usepackage[T1]{fontenc}
\usepackage{cuted}
\usepackage{siunitx}
\usepackage{caption}
\def\BibTeX{{\rm B\kern-.05em{\sc i\kern-.025em b}\kern-.08em
    T\kern-.1667em\lower.7ex\hbox{E}\kern-.125emX}}
\begin{document}

\title{Spatial IDFT for Squint-Free Massive Arrays\\
}

\author{\IEEEauthorblockN{Hesham Beshary, Ali Niknejad}
\IEEEauthorblockA{
Berkeley Wireless Research Center (BWRC), University of California, Berkeley, CA
}}

\maketitle

\begin{strip}
\centering
\fbox{%
  \parbox{0.97\textwidth}{%
    \footnotesize
    \textit{Accepted for publication in IEEE Transactions on Circuits and Systems I: Regular Papers.}

    \vspace{0.4em}

    \textcopyright\ 2026 IEEE. Personal use of this material is permitted.
    Permission from IEEE must be obtained for all other uses, in any current or future
    media, including reprinting/republishing this material for advertising or promotional
    purposes, creating new collective works, for resale or redistribution to servers or
    lists, or reuse of any copyrighted component of this work in other works.
  }%
}
\end{strip}

\begin{abstract}
This paper presents a novel technique to build squint-free massive phased arrays. This is accomplished by explicitly implementing a spatial IDFT to cancel out the DFT imposed by the array nature which causes beam squint. In addition, the paper analyzes the beam-squint issue, which arises from two mechanisms: the coherent bandwidth limitations and the systematic delay spread in the array. These mechanisms reduce the signal-to-noise ratio and cause intersymbol interference. This work also highlights the importance of utilizing OFDM modulation to enhance signal quality by mitigating the self-interference issue. A numerical solver is used to simulate and verify the IDFT squint-free implementation and to estimate the signal quality limitations in massive arrays.
\end{abstract}


\section{Introduction}


Phased array technology is favorable in various applications such as radar, wireless communication, and satellite systems for creating highly directional beams that can be steered electronically, enabling them to focus their signal energy in a specific direction. This energy focusing allows for increased range, enhanced signal-to-noise ratio (SNR), and interference reduction. Increasing the absolute bandwidth has been one of the primary solutions to address the ever-growing demand for data capacity in wireless systems. This leads to moving to mm-Wave/sub-THz frequency ranges for the bandwidth availability and higher absolute bandwidth for the same fractional one. These frequency ranges allow one to leverage phased arrays to overcome the free space path loss which is quadratically proportional to the operating frequency.

\begin{figure}[t]
\centering
\includegraphics[width=65mm]{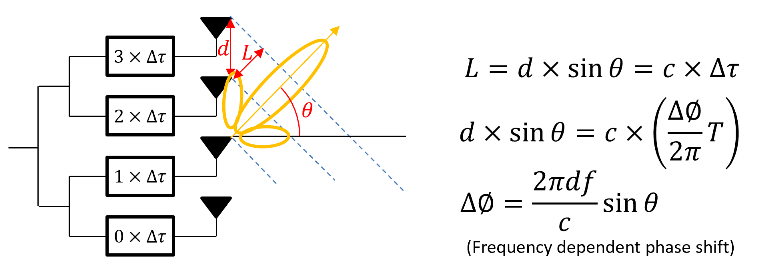}
\caption{Block diagram of a phased-array transmitter with true-time delay.}
\label{fig:intro_blk-diag}
\end{figure}

The directive beam synthesized by phased arrays can be electronically steered to a certain angle ($\theta_o$) by ideally generating progressive true-time delay (TTD) between the different elements ($\Delta \tau$) as shown in Fig. \ref{fig:intro_blk-diag}. TTD circuits generate variable group delay which is equal to the phase delay. mm-Wave/sub-THz TTD circuits can be implemented using variable relative delay lines \cite{ttd_1_1} or switching the signal flow among paths with different delay \cite{ttd_2_1,ttd_2_2}. However, these implementations usually suffer from large area consumption, limited range, and poor accuracy. Phase shifters achieve variable phase delay with zero relative group delay. In narrow-band systems, phase shifters \cite{ps_active,ps_pass1,ps_pass2,ps_pass3} offer good substitution for TTD circuits and are easier to implement, occupy compact area, and have unlimited range since phase wraps after $360^o$. However, the inequality between the group delay and the phase delay generated by phase shifters leads to beam squint in wideband systems. This issue becomes more apparent in massive arrays since the beam width is very narrow.

This paper focuses on the negative impact of beam squint on the transmitted and received signal bandwidth and quality in Section II. Section III shows that using orthogonal frequency-division multiplexing (OFDM) modulation is necessary in phased arrays to mitigate beam squint. A novel scheme for a squint-free massive array is introduced in Section IV. Finally, Section V concludes the paper.

\section{Array's Performance Degradation Due to Beam-Squint Issue}

\subsection{Coherent Bandwidth limitation}

\begin{figure}[t]
\centering
\includegraphics[width=65mm]{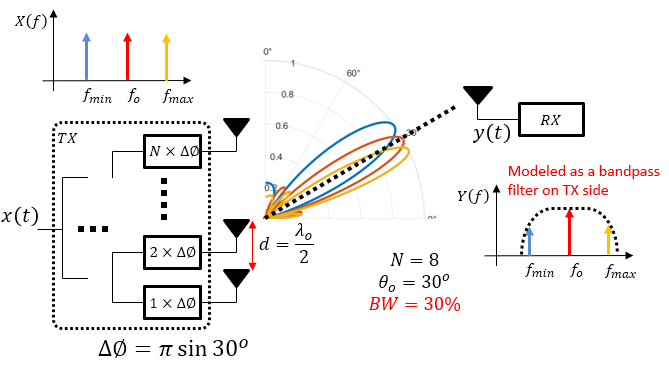}
\caption{Block diagram and space factor at the center frequency and band edges of a phased-array transmitter with phase shifters.}
\label{fig:perf_tx_blk-diag}
\end{figure}

Let us begin on the transmitter (TX) side. Phased arrays redistribute the power in space and boost the effective isotropic radiated power (EIRP) by $20\log{(N)}$ in specific direction, where $N$ is the number of array elements. As emphasized earlier, using phase shifters causes inequality between the phase delay and the group delay. This leads to different frequency points of the transmitted signal being steered to different angles. These points will not be received with the same magnitude at the receiver (RX), which will lead to EIRP degradation over frequency. Fig \ref{fig:perf_tx_blk-diag} shows an example of a linear phased array transmitter with $N=8$ elements, $30\%$ fractional bandwidth for the transmitted signal, and $d=\lambda_o/2$ spacing between each two neighboring elements, where $\lambda_o$ is the signal wavelength at the center frequency. The transmitted beam is steered to $\theta_o=30^o$ by generating progressive phase shifts between the elements, $\Delta\phi=2\pi(d/\lambda_o)\sin{\theta_o}=\pi\sin{30^o}$. The received signal experiences attenuation away from the center frequency.

In general, the normalized space factor magnitude of a linear array ($|SF_n|$) at any angle $\theta$ with respect to the array normal is given by
\begin{align*}
    |SF_n(\theta,\lambda)| & =\frac{1}{N}\left|\sum_{n=1}^{N}e^{j2\pi d(n-1)[(1/\lambda)\sin{\theta}-(1/\lambda_o)\sin{\theta_o}]}\right| \\
    &=\frac{1}{N} \left|  \frac{\sin{[N\pi d((1/\lambda)\sin{\theta}-(1/\lambda_o)\sin{\theta_o})]}}{\sin{[\pi d((1/\lambda)\sin{\theta}-(1/\lambda_o)\sin{\theta_o})]}}  \right|.
\end{align*}


Assuming that $d=\lambda_o/2$, which is reasonable to avoid grating lobes, then $d/\lambda=\lambda_o/2\lambda=f/2f_o$, where $f$ is the operating frequency and $f_o$ is the transmitted signal center frequency. At the steering direction ($\theta_o$), $|SF_n|$ is reduced to
$$
    |SF_n(\theta_o,f)| = \frac{1}{N} \left|  \frac{\sin{[N(\pi/2)\sin{\theta_o}([f/f_o]-1)]}}{\sin{[(\pi/2)\sin{\theta_o}([f/f_o]-1)]}}  \right|.   
$$

The array response can be modeled as a bandpass filter with certain fractional coherent bandwidth ($BW_c$). This bandwidth could be derived as follows. For $f=f_{3dB}=f_o(1+BW_c/2)$,
$$
    |SF_n(\theta_o,f_{3dB})|= \frac{1}{\sqrt{2}} = \frac{1}{N} \left|  \frac{\sin{[N(\pi/4) BW_c\sin{\theta_o}]}}{\sin{[(\pi/4) BW_c\sin{\theta_o}]}}  \right|.   
$$

For large arrays ($N\geq 8$), $|SF_n(\theta_o,f_{3dB})|$ approaches the following expression:
\begin{align*}
    |SF_n(\theta_o,f_{3dB})|= \frac{1}{\sqrt{2}} &\simeq \left|  \frac{\sin{[N(\pi/4) BW_c\sin{\theta_o}]}}{N[(\pi/4) BW_c\sin{\theta_o}]}  \right| \\
    & =|\text{sinc}[N(\pi/4) BW_c\sin{\theta_o}]|.
\end{align*}

Hence, the fractional coherent $3\ \text{dB}$ bandwidth is
$$
    BW_c \simeq \frac{1.77}{N \sin{\theta_o}}.
$$

As expected, the coherent bandwidth gets narrower as the array gets bigger, or the as the steering angle increases. As an example, if we assume 16 elements array with $30^o$ steering angle, the $BW_c$ is approximately equal to $22\%$. In addition, the array response experiences complete nulls due to fully-incoherent combining. The closest two null frequencies are,
$$
f_{null\pm}=f_o \pm \frac{2}{N \sin{\theta_o}},
$$
which are defined as $|SF_n(\theta_o,f_{null\pm})|=0$.

By reciprocity, the limited coherent bandwidth issue appears on the RX side and can be explained with the same analysis. However, the action of combining happens electrically inside the array, unlike the TX which happens over the air. This limited bandwidth issue leads to SNR degradation on the RX side, which will be discussed later in Section III.

\subsection{Systematic Delay Spread and Inter-Symbol Interference}

\begin{figure}[t]
\centering
\includegraphics[width=80mm]{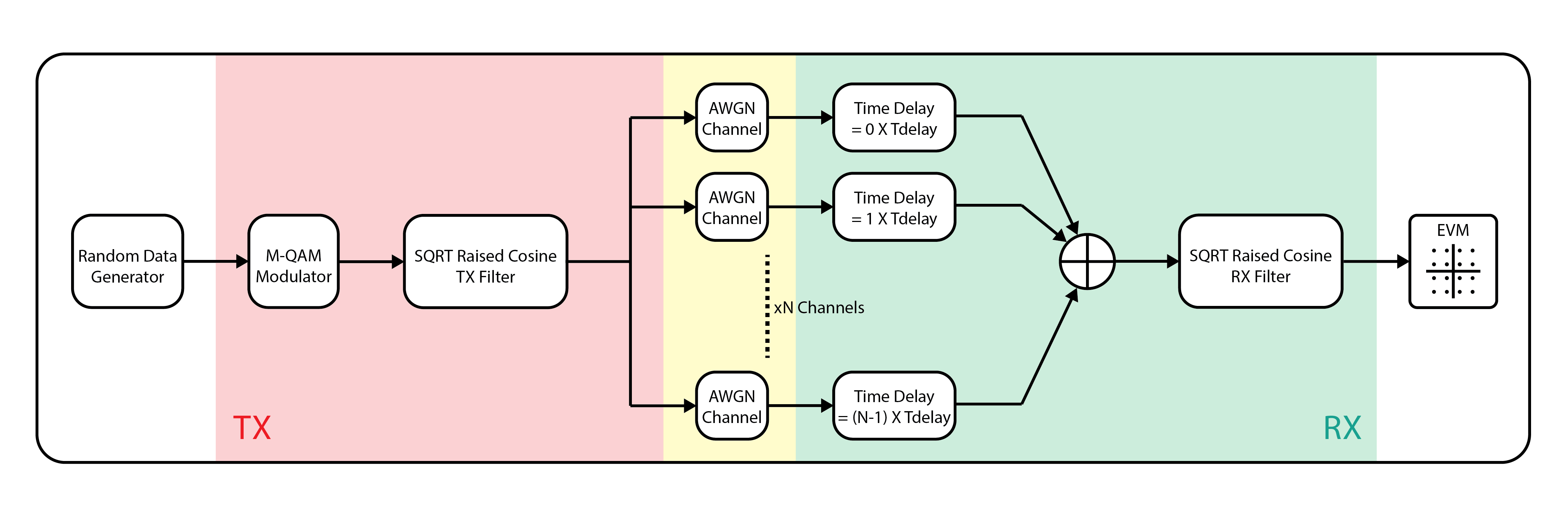}
\caption{Baseband model of a transceiver with single TX and N-elements phased array RX using phase shifters and processing single-carrier signal.}
\label{fig:perf_rx_blk-diag}
\end{figure}

\begin{figure}[t]
\centering%
\subfloat[]{%
\centering
\includegraphics[width=45mm]{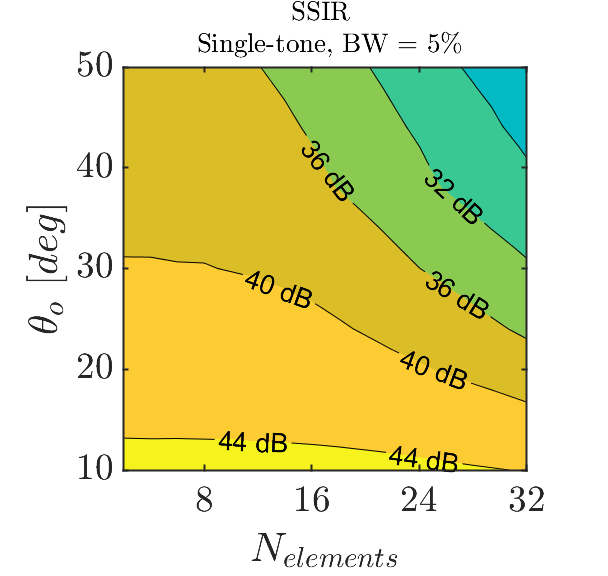}
\label{fig:rx_SSIR_ST_out_5}
}
\subfloat[]{%
\centering
\includegraphics[width=45mm]{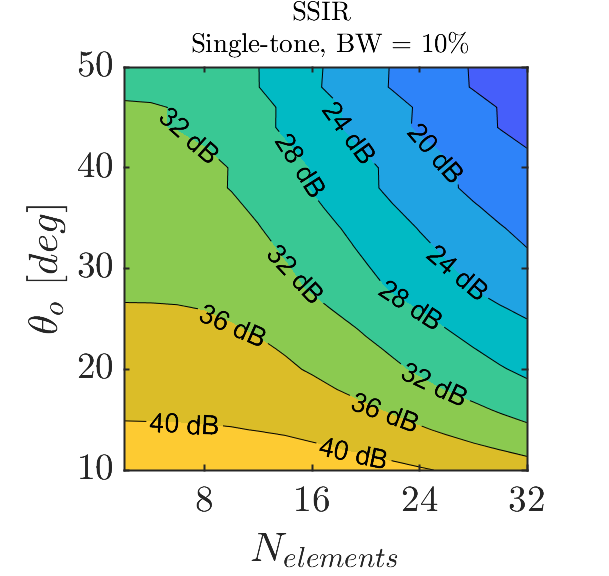}
\label{fig:rx_SSIR_ST_out_10}
}%

\subfloat[]{%
\centering
\includegraphics[width=45mm]{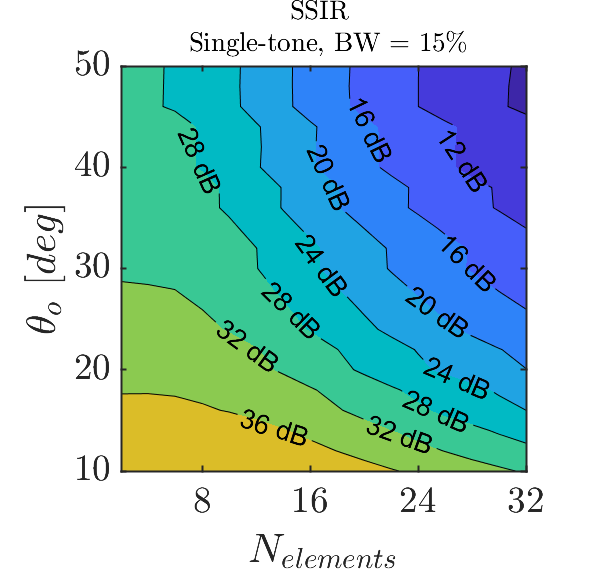}
\label{fig:rx_SSIR_ST_out_15}
}
\subfloat[]{%
\centering
\includegraphics[width=45mm]{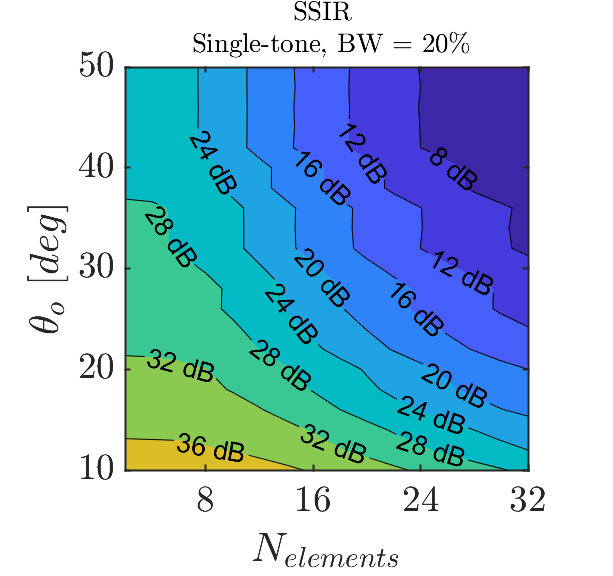}
\label{fig:rx_SSIR_ST_out_20}
}%
\\[2mm]
\caption{SSIR over array size $N_{elements}$ and steering angle $\theta_o$ assuming a wideband single-carrier signal for (a) $5\%$, (b) $10\%$, (c) $15\%$, and (d) $20\%$ fractional bandwidth}
\label{fig:rx_SSIR_ST_out}
\end{figure}

\begin{figure}
\centering%
\subfloat[]{%
\centering
\includegraphics[width=43mm]{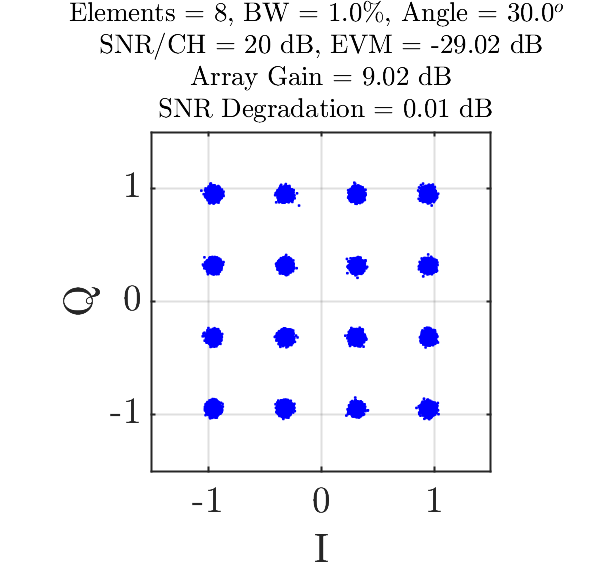}
\label{fig:rx_cont_1}
}
\subfloat[]{%
\centering
\includegraphics[width=43mm]{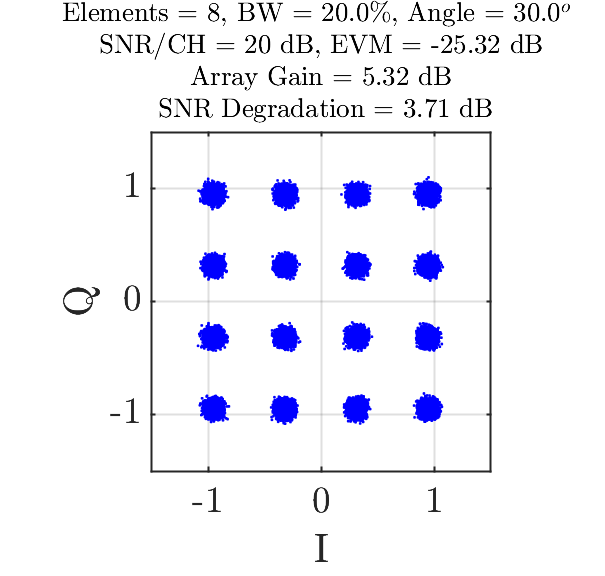}
\label{fig:rx_cont_2}
}%

\subfloat[]{%
\centering
\includegraphics[width=43mm]{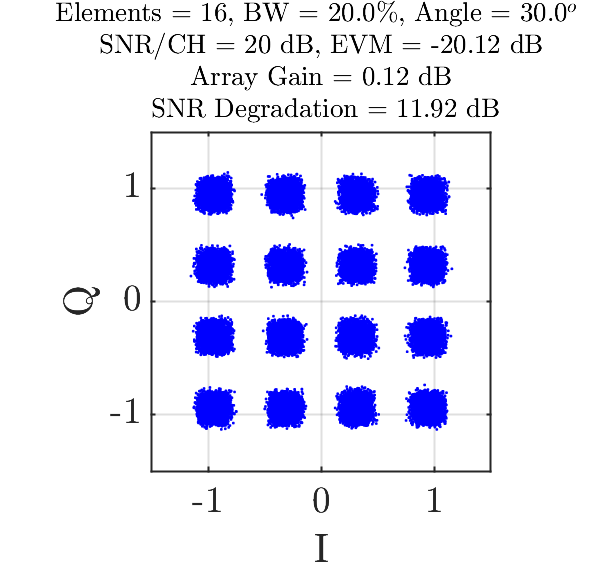}
\label{fig:rx_cont_3}
}
\subfloat[]{%
\centering
\includegraphics[width=43mm]{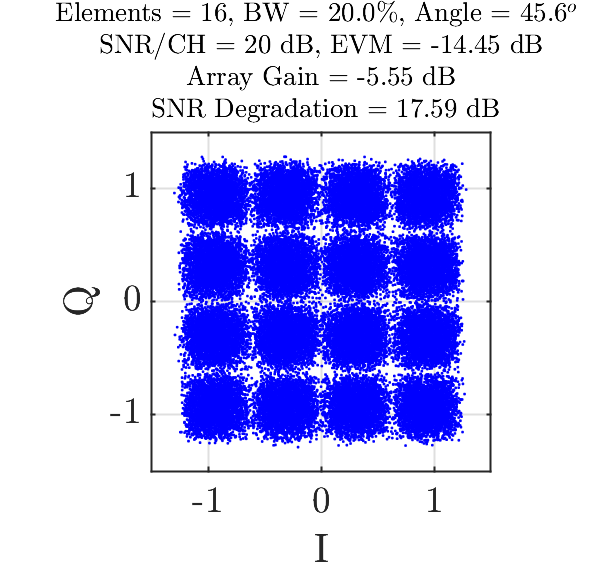}
\label{fig:rx_cont_4}
}%
\\[2mm]
\caption{Constellation and EVM simulation results for a single-carrier signal with $20\ \text{dB}$ SNR received by an 8-elements RX at $30^o$ steering angle and (a) $1\%$ bandwidth (b) $20\%$ bandwidth, and a 16-elements RX with $20\%$ bandwidth and (c) $30^o$ steering angle (d) $45^o$ steering angle }
\end{figure}

On the RX side, phased arrays lead to SNR enhancement by $10\log{(N)}$ since they coherently combine the signal from different channels and the noise is incoherently combined. While receiving from a non-zero angle with respect to the array normal, the signal is received with progressive time delay at the different elements of the array. Phase shifters are able to correct for the different phase delay to align the center frequency for coherent combining, however, the group delay will not be affected. Hence, the baseband signals from the different channels will still contain progressive time delay. This uncorrected delay causes systematic delay spread ($T_{spread}$) in the combined signal which leads to intersymbol interference (ISI).

The hard limit to avoid this ISI issue is having the symbol period ($T_{symbol}$) greater than the maximum delay spread, which will occur between the first element and the last one. Therefore,
$$
T_{symbol} > T_{spread-max},
$$
$$
T_{symbol} > N \times \Delta\tau,
$$
$$
T_{symbol} > \frac{Nd}{c} \sin{\theta_o},
$$
where $c$ is the free-space propagation speed of the signal, which is equal to $c=\lambda_o f_o$. The passband signal bandwidth is roughly $1/T_{symbol}$. Hence, the signal fractional bandwidth is $BW_{sig} = 1/(f_o T_{symbol})$. Therefore,
$$
\frac{1}{BW_{sig} f_o } > \frac{N(\lambda_o/2)}{\lambda_o f_o} \sin{\theta_o},
$$
$$
BW_{sig} < \frac{2}{N \sin{\theta_o}}.
$$

As expected, the final result is compatible with the $BW_c$ that is derived on the TX side. Practically, the amount of ISI and SNR degradation depends on the transceiver architecture, such as pulse shaping, which complicates the analytical analysis. This issue can easily be analyzed using a numeric computing environment. Fig. \ref{fig:perf_rx_blk-diag} shows a block diagram of a phased array receiver using MATLAB to simulate the error vector magnitude (EVM) and signal-to-self-interference ratio (SSIR) for different scenarios. In this simulation, SSIR is defined as the modulation error ratio (MER) assuming a noiseless ideal channel. The simulated SSIR versus the number of elements ($N$) and the steering angle ($\theta_o$) is shown in Fig. \ref{fig:rx_SSIR_ST_out}\subref*{fig:rx_SSIR_ST_out_5}--\subref*{fig:rx_SSIR_ST_out_20} for $5\%$, $10\%$, $15\%$ and $20\%$ fractional bandwidth, respectively. This basically shows the upper limit of the signal quality. In presence of additive white Gaussian noise (AWGN) channel, the overall EVM can be calculated as follows:
$$
\text{EVM}=\frac{1}{\text{MER}} =\sqrt{\frac{1}{\text{SNR}^2} + \frac{1}{\text{SSIR}^2}}.
$$

It is instructive to compare the input signal SNR at each channel with the array input-referred SSIR, which is defined as
$$
\text{SSIR}_{in}=\text{SSIR}-10\log(N).
$$
If the input SNR per channel is larger than the $\text{SSIR}_{in}$, EVM will be dominated by the array SSIR and increasing $N$ will reduce the signal quality, worsen the EVM, and increase the system power consumption.

Fig. \ref{fig:rx_cont_1} shows the simulation results of 8-elements receiver with $1\%$ fractional bandwidth, $30^o$ receiving angle and 20 dB input SNR. This narrowband system does not suffer from self interference since $\text{SSIR}_{in}$ is significantly larger than the input SNR. As expected, the output signal quality is improved compared to the input by the array gain, which is equal to $10\log{(N)}=9 \ \text{dB}$, and $\text{EVM}=-29$ dB. The beam-squint issue is apparent in this system with $20\%$ bandwidth signal since SNR and SSIR are in the same order. EVM and the effective array gain degrade by $\approx3.7 \ \text{dB}$ as shown in Fig. \ref{fig:rx_cont_2}. Increasing the number of elements and the receiving angle worsen the EVM, as shown in Fig. \ref{fig:rx_cont_3} and Fig. \ref{fig:rx_cont_4}, since EVM is mainly dominated by SSIR. This self-interference issue is the dominant performance degradation mechanism in phased arrays with single-carrier wideband signal. This issue also appears on the TX side for the over-the-air combined signal.

\section{OFDM for Self Interference Mitigation}

\begin{figure}[t]
\centering
\includegraphics[width=88mm]{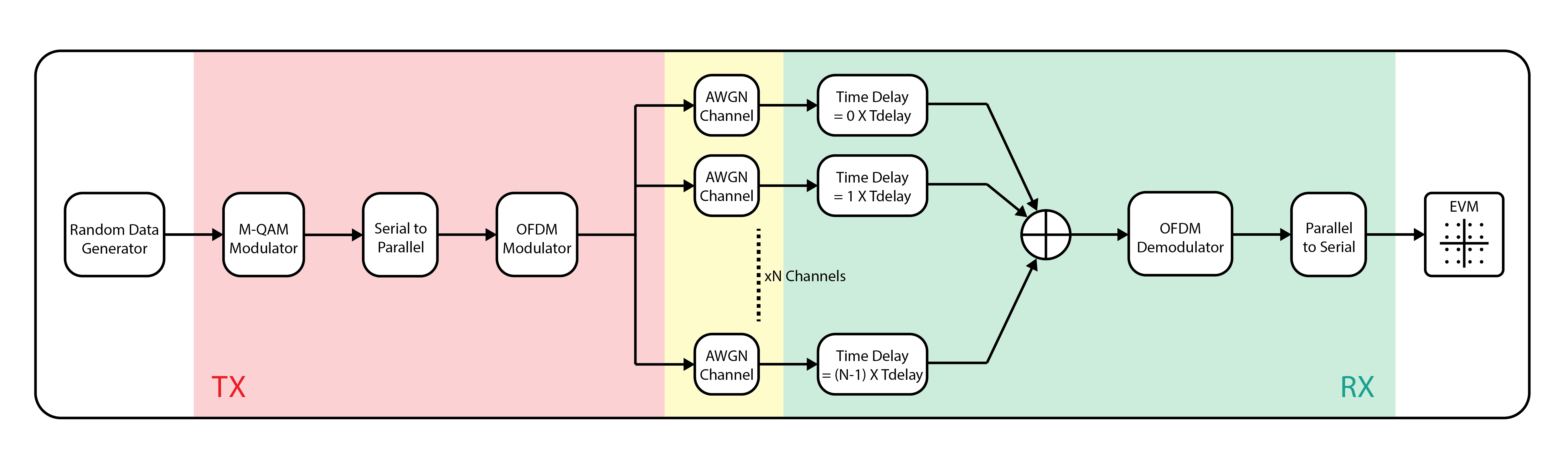}
\caption{Baseband model of a transceiver with single TX and N-elements phased array RX using phase shifters and processing an OFDM signal.}
\label{fig:ofdm_blk-diag}
\end{figure}

\begin{figure}[t]
\centering%
\subfloat[]{%
\centering
\includegraphics[width=45mm]{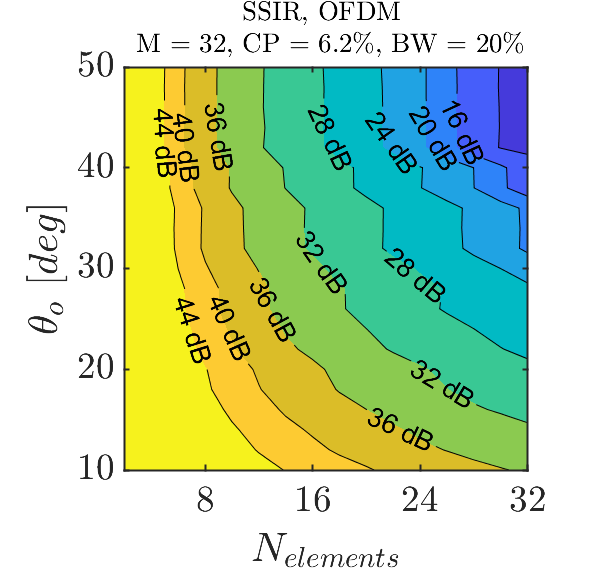}
\label{fig:rx_SSIR_OFDM_out_20_32}
}%
\subfloat[]{%
\centering
\includegraphics[width=45mm]{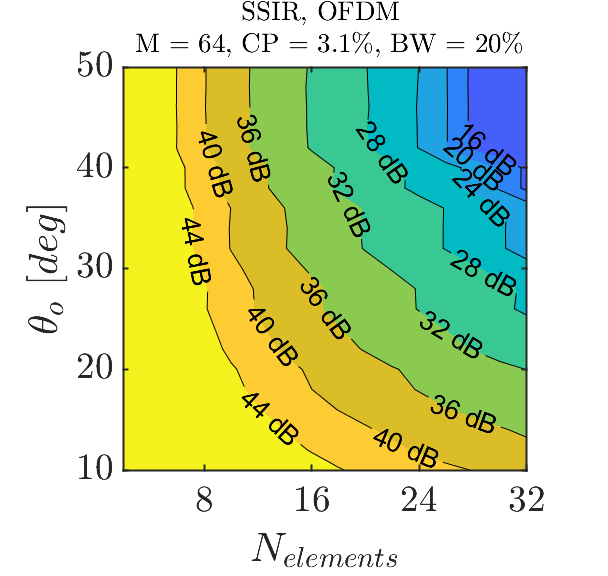}
\label{fig:rx_SSIR_OFDM_out_20_64}
}%
\\[2mm]
\caption{SSIR over array size $N_{elements}$ and steering angle $\theta_o$ for a $20\%$ OFDM signal and (a) 32 subcarriers, and (b) 64 subcarriers}
\end{figure}

\begin{figure}[t]
\centering%
\subfloat[]{%
\centering
\includegraphics[width=50mm]{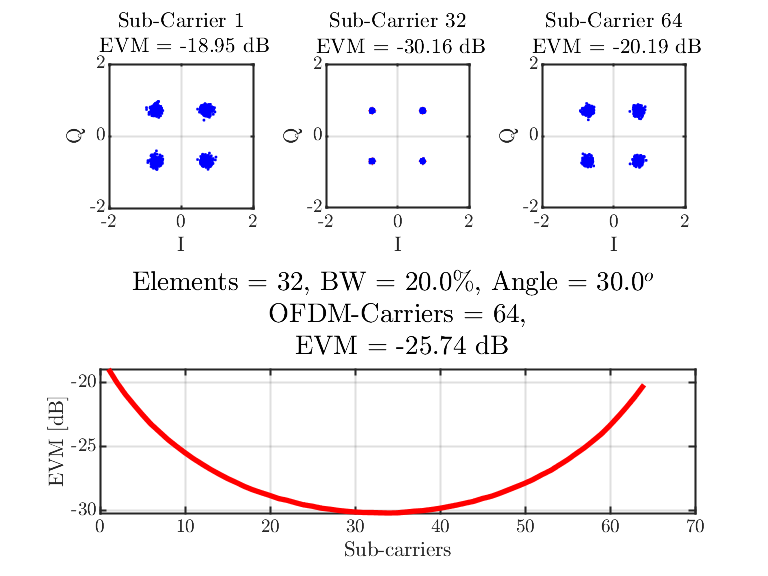}
\label{fig:rx_OFDM_1_carr64}
}%
\subfloat[]{%
\centering
\includegraphics[width=38mm]{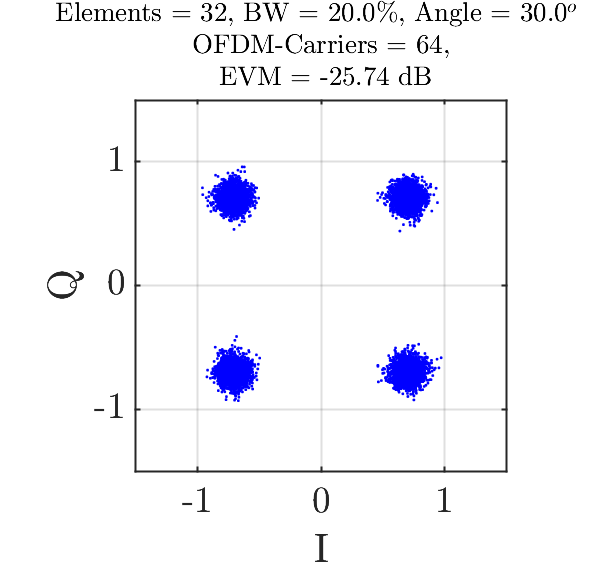}
\label{fig:rx_OFDM_2_carr64}
}%
\\[2mm]
\caption{(a) EVM over different OFDM tones for 32-elements RX, 64 subcarriers, $20\%$ signal bandwidth, $30^o$ receiving angle and noiseless ideal channel, and (b) overall signal constellation and EVM}
\end{figure}

\begin{figure}[t]
\centering%
\subfloat[]{%
\centering
\includegraphics[width=50mm]{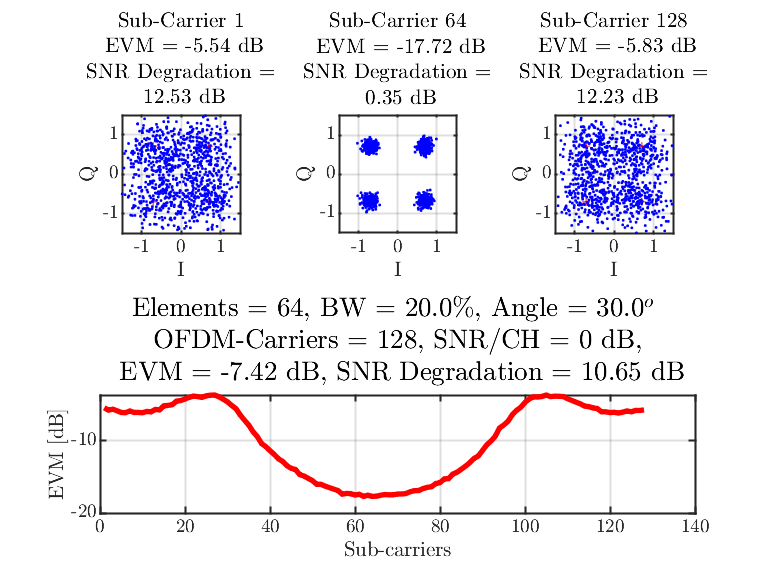}
\label{fig:rx_OFDM_1_carr128}
}%
\subfloat[]{%
\centering
\includegraphics[width=37mm]{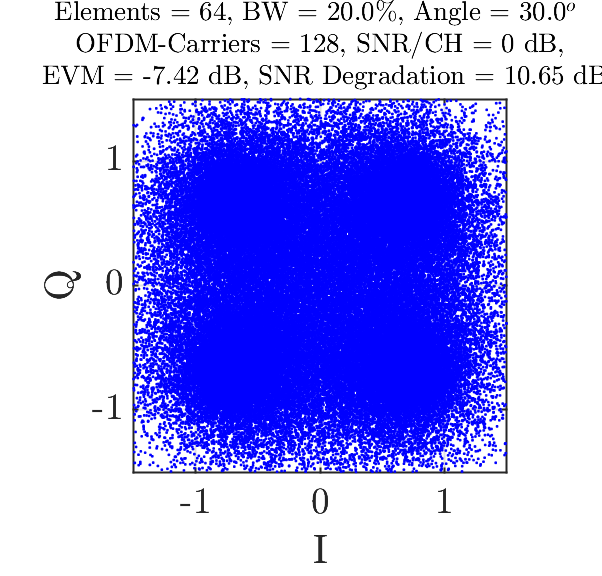}
\label{fig:rx_OFDM_2_carr128}
}%
\\[2mm]
\caption{(a) EVM over different OFDM tones for 64-elements RX, 128 subcarriers, $20\%$ signal bandwidth, $30^o$ receiving angle and $0\ \text{dB}$ input SNR, and (b) overall signal constellation and EVM}
\label{fig:rx_OFDM_carr128}
\end{figure}

OFDM significantly reduces the self interference in phased arrays since it is robust against ISI and has much higher tolerance against the systematic delay spread. The hard limit for ISI will be relaxed to
$$
T_{OFDM-symbol} > T_\text{spread-max},
$$
$$
T_{symbol} > \frac{Nd}{Mc} \sin{\theta_o},
$$
where $M$ is the number of OFDM subcarriers. However, the cyclic prefix ($T_{cp}$) should be larger than $T_\text{spread-max}$ to keep orthogonality between different subcarriers and avoid inter-carrier interference. It also reduces the residue ISI in the signal. Hence, minimum value for $T_{cp}/T_{symbol}$ is $1/M$. The improvements of using OFDM modulation in phased arrays can similarly be simulated by implementing the system in Fig. \ref{fig:ofdm_blk-diag} using MATLAB. For $20\%$ fractional bandwidth, Fig. \ref{fig:rx_SSIR_OFDM_out_20_32} and Fig. \ref{fig:rx_SSIR_OFDM_out_20_64} show the simulated SSIR for 32 and 64 OFDM subcarriers, respectively, and $T_{cp}/T_{symbol}$ equal to $2/M$. Compared to a single-carrier signal, OFDM reduced self interference by $\approx 15\ \text{dB}$.

Even though OFDM mitigates the ISI due to the systematic delay spread, it still suffers from the limited coherent bandwidth issue. This can be observed by looking at the EVM vs different OFDM bins as shown in Fig. \ref{fig:rx_OFDM_1_carr64} for 32 elements, $20\%$ fractional bandwidth, $30^o$ steering angle, 64 OFDM subcarriers, and noiseless ideal channel where EVM is dominated by SSIR. EVM is minimum at the center frequency, which occurs at the center tone, and starts to degrade as the frequency moves to the band edges. This eventually degrades EVM from $-29.8\ \text{dB}$ for only the center tone to $-25.4\ \text{dB}$ for the overall transmitted data as shown in Fig. \ref{fig:rx_OFDM_2_carr64}. Deep fading can occur when the OFDM bin suffers from complete incoherent combining as shown in Fig \ref{fig:rx_OFDM_1_carr128} when the number of elements is doubled to $64$, subcarriers are also doubled to keep the self interference level almost the same at the center frequency. A $20\ \text{dB}$ input SNR is considered for this simulation. The deep-faded bins can be calculated from the null frequencies in $|SF_n|$ ($f_{null\pm}$) as follows:
$$
M_{null\pm}=M_{carriers}\times \left( \frac{1}{2} \pm \frac{2}{N\times BW_{sig} \times \sin{\theta_o}} \right).
$$
Similarly, the maximum number of tones around the center frequency that allow for $3 \ \text{dB}$ SNR degradation can be calculated from the coherent bandwidth relation as following:
$$
M_{3\text{dB}}=M_{carriers}\times \frac{1.77}{N\times BW_{sig} \times \sin{\theta_o}}.
$$

\section{Spatial IDFT for Fully-Coherent Squint-Free Massive Phased Arrays}

\begin{figure}[t]
\centering%
\subfloat[]{%
\centering
\includegraphics[width=80mm]{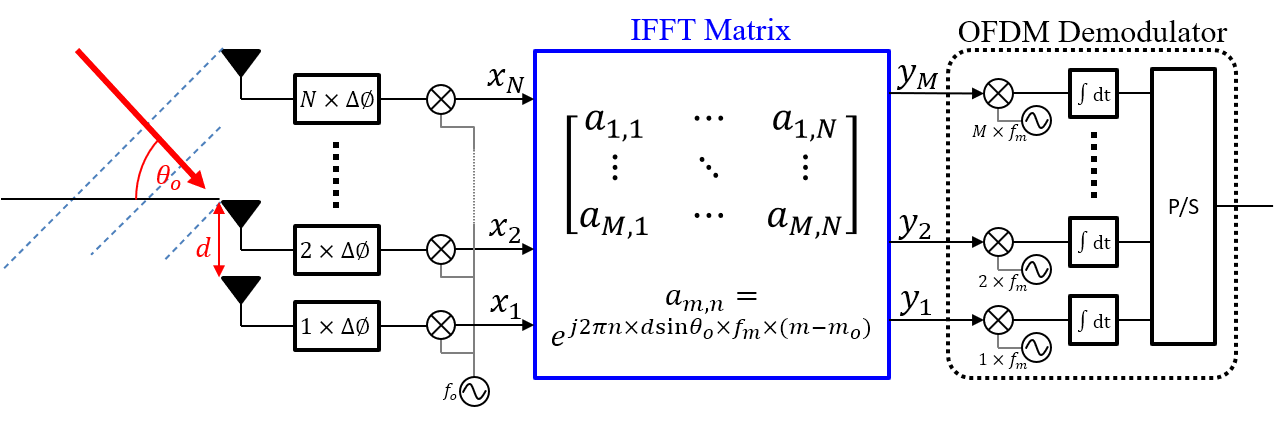}
\label{fig:rx_blk_diag_OFDM_DFT_Matrix}
}%

\subfloat[]{%
\centering
\includegraphics[width=85mm]{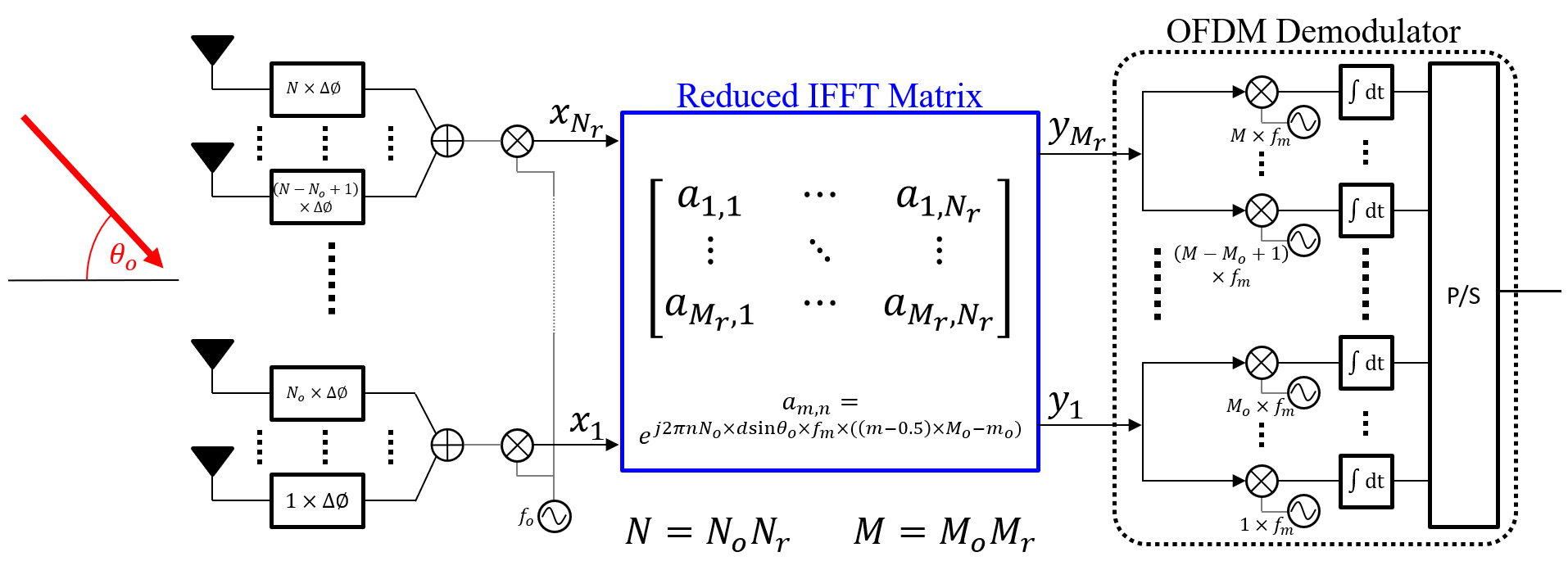}
\label{fig:rx_blk_diag_OFDM_reducedDFT_Matrix}
}%
\\[2mm]
\caption{Proposed architecture for squint-free massive arrays using (a) full IDFT matrix and (b) reduced IDFT matrix.}
\end{figure}

\begin{figure}[t]
\centering
\includegraphics[width=88mm]{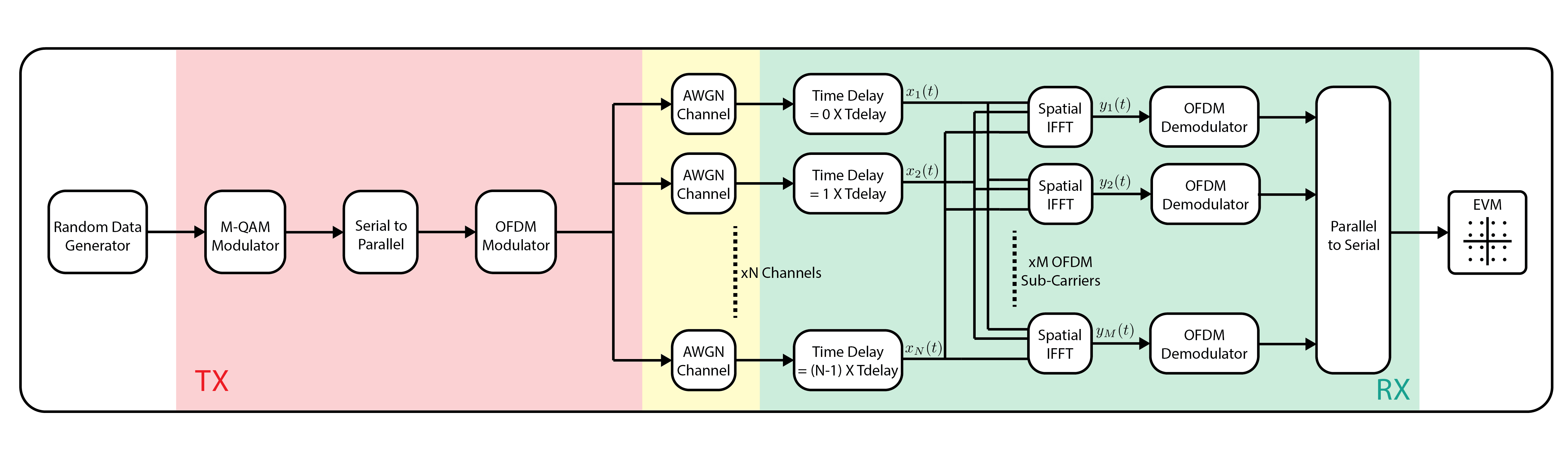}
\caption{Baseband model of a transceiver with single TX and N-elements phased array RX using phase shifters and full IDFT matrix implementation}
\label{fig:ofdm_DFT_blk-diag}
\end{figure}

Phased arrays perform Discrete Fourier transform (DFT) on the processed signal. This DFT is the source of the beam-squint issues in phased arrays. Inverse DFT (IDFT) can electrically be implemented in the back-end to cancel the DFT effect. On the RX side, this can be done, as shown in Fig. \ref{fig:rx_blk_diag_OFDM_DFT_Matrix}, by processing the signals received from different elements $x_n(t)$ as follows:
$$
y_m(t)=\sum_{n=1}^{N} x_n(t) \times e^{j2\pi n \times d \sin{\theta_o} \times \Delta f_m \times (m-m_o)}
$$
where $y_m(t)$ signals are the inputs to the OFDM demodulator, $\Delta f_m$ is the subcarrier spacing, and $m_o$ is the tone number that falls at the center frequency ($f_o$) in passband. Quantitatively, IDFT creates a stair-case like phase response for different OFDM tones. This phase response has a virtual group-delay effect that realigns the OFDM signal from different channels, which allows for coherent combining of all the tones. 

\begin{figure}[t]
\centering%
\subfloat[]{%
\centering
\includegraphics[width=50mm]{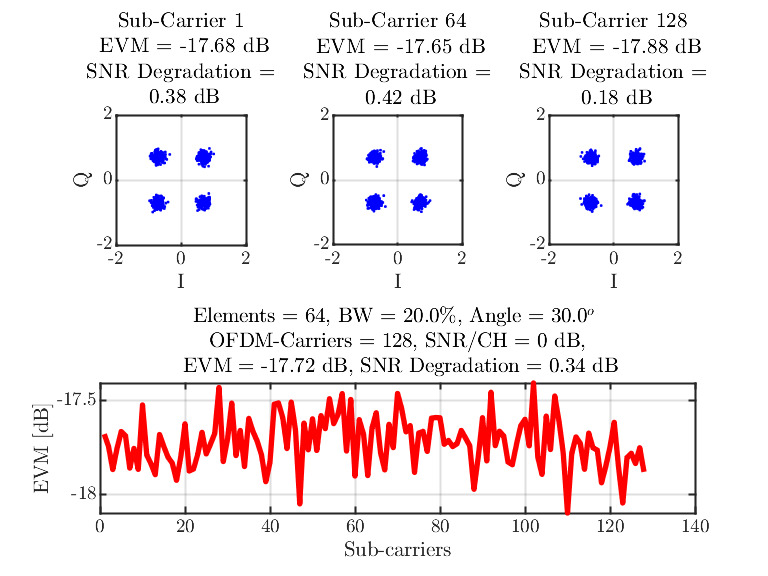}
\label{fig:rx_OFDM_IDFT_1_carr128}
}%
\subfloat[]{%
\centering
\includegraphics[width=37mm]{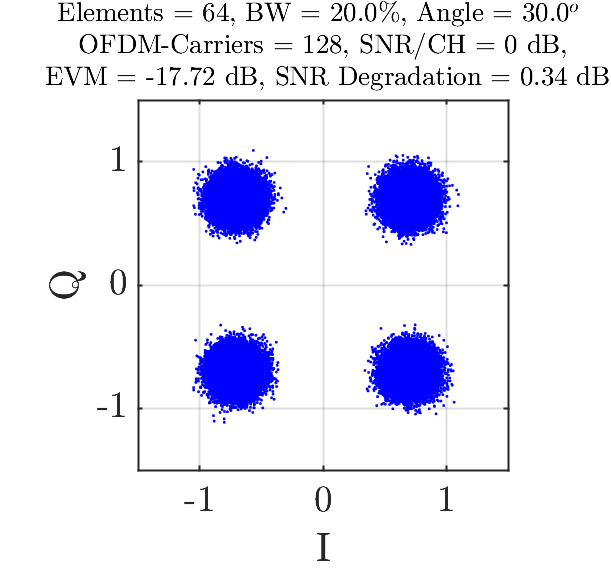}
\label{fig:rx_OFDM_IDFT_2_carr128}
}%
\\[2mm]
\caption{(a) EVM over different OFDM tones using the proposed full IDFT system for 64-elements RX, 128 subcarriers, $20\%$ signal bandwidth, $30^o$ receiving angle and $0\ \text{dB}$ input SNR, and (b) overall signal constellation and EVM}
\label{fig:rx_OFDM_IDFT_carr128}
\end{figure}

Similarly, the proposed system is verified and simulated using MATLAB by implementing the system in Fig. \ref{fig:ofdm_DFT_blk-diag}.  To apply a one-to-one comparison, the same system and signal parameters used in Fig. \ref{fig:rx_OFDM_carr128} are simulated through the proposed IDFT receiver. The received signal quality remains the same and the EVM is almost constant for all the subcarriers as shown in Fig. \ref{fig:rx_OFDM_IDFT_1_carr128}. In this example, the IDFT matrix dimension is $M \times N = 128 \times 64$. Practically, this system is very complicated and hard to implement. 

\begin{figure}[t]
\centering%
\subfloat[]{%
\centering
\includegraphics[width=50mm]{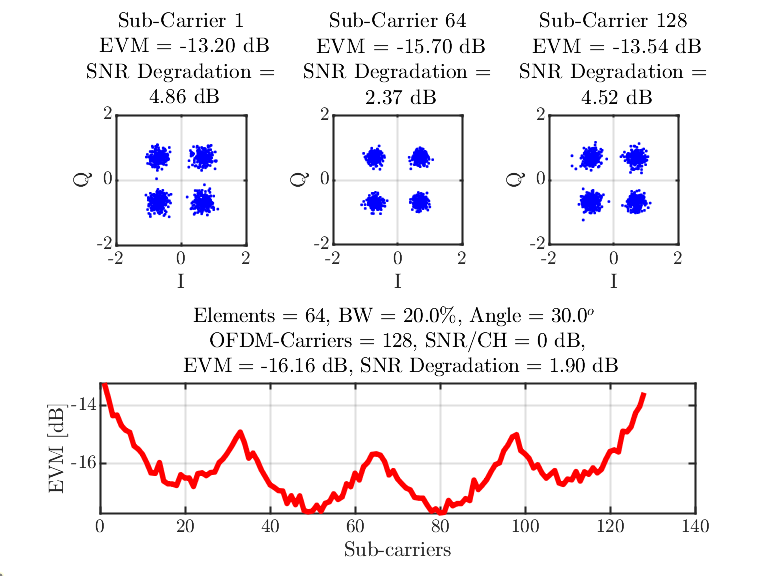}
\label{fig:rx_OFDM_reducedIDFT_1_carr128}
}%
\subfloat[]{%
\centering
\includegraphics[width=37mm]{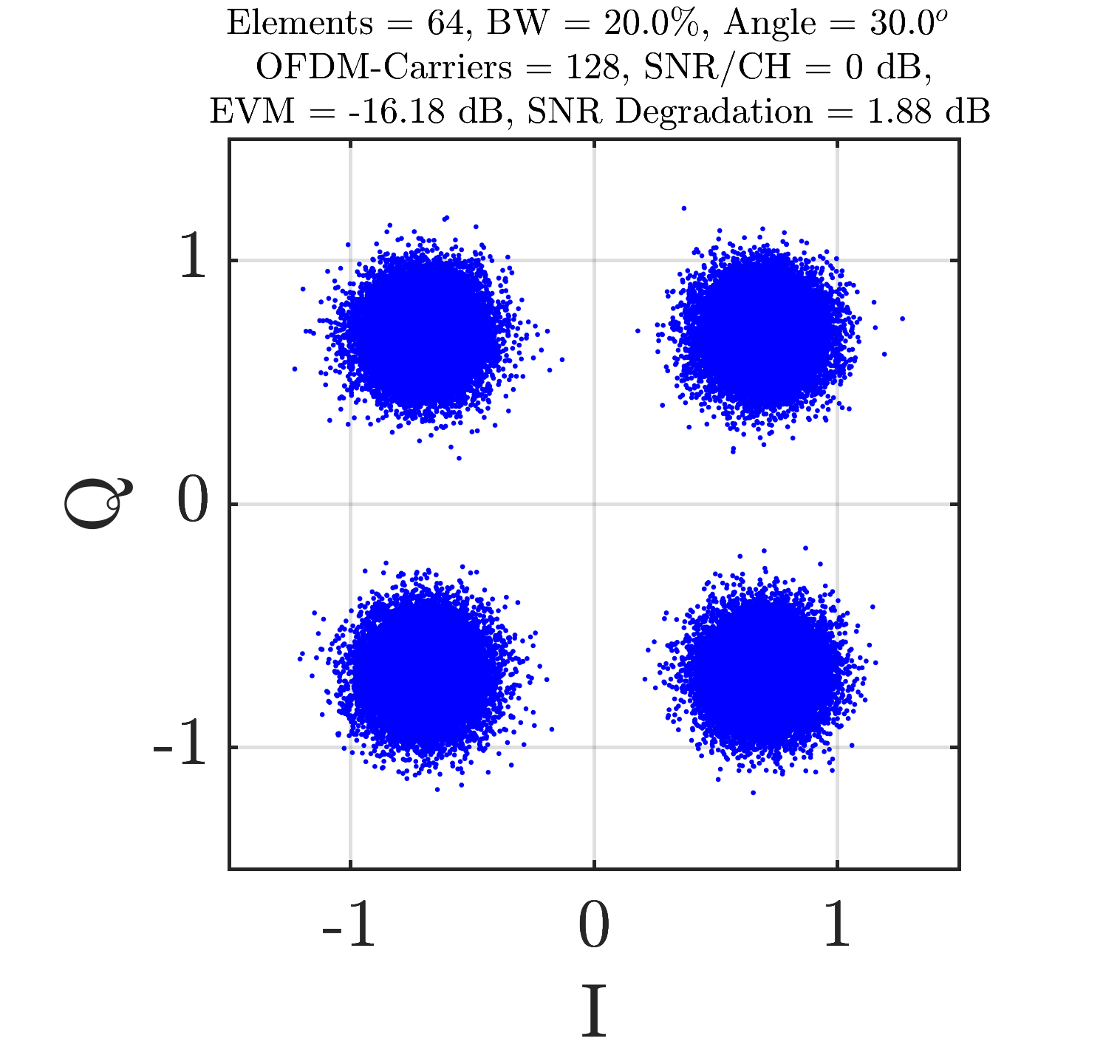}
\label{fig:rx_OFDM_reducedIDFT_2_carr128}
}%
\\[2mm]
\caption{(a) EVM over different OFDM tones using the proposed reduced IDFT system for 64-elements RX, 128 subcarriers, $20\%$ signal bandwidth, $30^o$ receiving angle and $0\ \text{dB}$ input SNR, and (b) overall signal constellation and EVM}
\end{figure}

A reduced IDFT matrix is more practical which allows for some beam squint and less degradation in the signal quality to simplify the system and reduce the IDFT matrix size. The reduced solution is shown in Fig. \ref{fig:rx_blk_diag_OFDM_reducedDFT_Matrix} and can be implemented by pre-combining sub-arrays with size $N_o$, where $N=N_o \times N_r$. Accepting $3\ \text{dB}$ degradation, the coherent bandwidth of the sub-array must be larger than the signal bandwidth. Each output of the IDFT matrix is used to demodulate $M_o$ tones, where $M=M_o \times M_r$. For $3\ \text{dB}$ EVM ripples, $M_o$ must be smaller than $M_{3\text{dB}}$. Therefore, 
$$
N_o<\frac{1.77}{BW_{sig}\times\sin{\theta_{o}}},
$$
and
$$
M_o<M\times \frac{1.77}{N\times BW_{sig} \times \sin{\theta_o}}.
$$
Hence, the IDFT matrix size is reduced to $M_r \times N_r$, where
$$
(N_r,M_r)> \frac{N\times BW_{sig} \times \sin{\theta_o}}{1.77}.
$$
For the simulated system in Fig. \ref{fig:rx_OFDM_IDFT_carr128}, $M_r \times N_r$ can be calculated to be $4 \times 4$, which is a significant reduction and more practical compared to the full matrix size. The same system and signal parameters are simulated using the reduced IDFT architecture and the results are shown in Fig. \ref{fig:rx_OFDM_reducedIDFT_1_carr128}, which shows $\approx 3\ \text{dB}$ fluctuations for the EVM as designed.

\section{Conclusion}
In this work, the beam-squint issue is carefully analyzed to relate the array parameters to the coherence bandwidth and the systematic delay spread. The signal self-interference due to this delay spread is simulated under specific assumptions, and is shown to be the limiting factor to the signal quality in phased arrays. Finally, squint-free massive arrays can be built using a spatial IDFT, which is applied on an OFDM signal. 

\section*{Acknowledgment}
This material is based upon work supported by the National Science Foundation under Grant No. EECS-2148021

\small{
\bibliographystyle{IEEEtran}

\bibliography{IEEEexample}
}

\end{document}